\documentclass{DISproc}

\begin{document}
\title{Inelastic proton-proton cross section measurements in CMS at $\sqrt{s}=7$ TeV}

\author{{\slshape Anna Julia Zsigmond$^{1,2}$ on behalf of the CMS Collaboration}\\[1ex]
$^1$E\"otv\"os Lor\'and University, Budapest, Hungary\\
$^2$Institute for Particle and Nuclear Physics, Wigner RCP, Budapest, Hungary }

\contribID{181}

\doi  

\maketitle

\begin{abstract}
We present measurements of the total inelastic pp cross section at \unit{7}{\TeV} obtained with the CMS detector. Two different methods are used. In runs with low event pile-up, we exploit the large pseudorapidity coverage ($|\eta|<5.2$) of the CMS calorimeters to obtain the cross section for events with any activity in the acceptance range. In addition, runs with high event pile-up are used by fitting a Poisson distribution with the total visible cross section as parameter to the number of reconstructed primary vertices. Both measurements are corrected to a hadron level definition of the inelastic cross section.
\end{abstract}

\section{Introduction}

The cross sections of hadronic collisions are important and fundamental quantities in high energy particle and nuclear physics and have been studied in the last 40 years in experiments covering many orders of magnitude in center-of-mass energies.

In this report we present two methods for measuring the inelastic proton-proton cross section with the CMS detector. Data collected in 2010 were used in both analyses but with different luminosity and pile-up conditions. Low pile-up data were used to count events with activity in either of the Hadron Forward Calorimeters (HF) in the first method presented. The other method used high luminosity data to count pile-up events in a given bunch crossing as a function of bunch luminosity and fitted with a Poisson distribution to evaluate the cross section.

Both measurements are corrected to hadron level definitions of the inelastic $pp$ cross section. From these definitions, extrapolations are performed to the total inelastic $pp$ cross section with various Monte Carlo models also used in cosmic-rays physics.

\section{Event counting method with single-sided trigger}

The goal of the first method~\cite{CMSQCD} is to count events with as loose selection as possible to detect the largest possible cross section which translates to the smallest possible extrapolation. In the event counting, we required one reconstructed energy deposit in either side of the HF Calorimeters with at least 5~GeV total energy. The possible signal events were triggered by two proton bunches entering CMS and the background was estimated from the number of selected events triggered by a single bunch entering CMS. With the HF detectors only, more than one inelastic interaction in the same bunch- crossing cannot be separated, which means the need of low pile-up data ($\lambda<12\%$) for small pile-up correction factor.

Using generator level Monte Carlo, the inelastic interactions can be characterized with the variable $\xi$. The final state particles are ordered in rapidity and the largest rapidity gap is used to assign the particles into two systems. The invariant masses of the two systems are calculated and the higher mass system is called $X$, then the variable $\xi$ is given by $\xi = M_X^2/s$.
In case of single diffractive events, $\xi$ is the fractional momentum loss of the scattered proton.
Studying the HF selection efficiency as a function of $\xi$ showed that events with small $\xi$ can escape detection. For $\xi>5\times10^{-6}$, the efficiency of detection is more than $~98\%$.

The definition of the inelastic $pp$ cross section with $\xi>5\times10^{-6}$ is
\[
\sigma_{\rm{inel}}(\xi > 5\times10^{-6}) = \frac{N_{\rm{inel}}(1-f_{\xi})F_{\rm{pile-up}}}{\epsilon_{\xi}\int\mathcal{L}\rm{d}t}
\]
where $N_{\rm{inel}}$ is the number of events selected by the HF calorimeters after subtracting the background, $f_{\xi}$ is the fraction of selected events that are low mass (contamination), $F_{\rm{pile-up}}$ is the pile-up correction factor, $\int\mathcal{L}\rm{d}t$ is the integrated luminosity and $\epsilon_{\xi}$ is the efficiency to detect an inelastic event with $\xi>5\times10^{-6}$, namely the fraction of high mass events that are selected.

The efficiency ($\epsilon_\xi$) and contamination ($f_\xi$) correction factors were determined from three Monte Carlo generators using the full detector simulation of CMS: {\sc pythia 6}~\cite{Pythia6}, {\sc pythia 8}~\cite{Pythia8} and {\sc phojet}~\cite{Phojet}. To calculate the pile-up correction factor ($F_{\rm{pile-up}}$) an iterative method was used where the average number of collisions per bunch-crossing, the pile-up ($\lambda$) was measured from the data directly. Several low pile-up datasets were used to obtain the cross section. The integrated luminosity values for these datasets were obtained on the basis of Van der Meer scans, which carry a 4\% normalization uncertainty dominating the uncertainties in the present analyses.

The result of the analysis for events with $(\xi>5\times10^{-6})$ is obtained by averaging the 5~GeV HF threshold cross section values measured at different pile-up conditions. The systematic uncertainties of the result take into account noisy tower exclusion, run-by-run luminosity variations, varying the HF energy threshold and the model dependence.

The final result for the inelastic $pp$ cross section with $(\xi>5\times10^{-6})$:
\[ 
\sigma_{\rm{inel}}(\xi>5\times10^{-6})=60.2\pm0.2(\rm{stat.})\pm1.1(\rm{syst.})\pm2.4(\rm{lumi.})\mbox{ mb.} 
\]

\section{Pile-up counting method}

The second method~\cite{CMSFWD} is based on the assumption that the number of inelastic $pp$ interactions in a given bunch crossing follows a Poisson probability distribution:
\[
 P(n) = \frac{(\mathcal{L}\sigma_{\rm{inel}})^{n}}{n!}e^{-\mathcal{L}\sigma_{\rm{inel}}}
\]
where $\mathcal{L}$ is the bunch crossing luminosity and $\sigma_{\rm{inel}}$ is the total inelastic $pp$ cross section.

Two data samples, an inclusive di-electron and a single muon, were collected with the CMS triggers. The specific trigger requirements are not important as long as their efficiencies do not depend on the instantaneous luminosity. Using these data samples the number of reconstructed vertices were counted in each bunch crossing in bins of luminosity. Every vertex had to fulfill the quality requirements to be counted as a visible vertex. The quality requirements are the following: the transverse position of the vertex between $\pm0.06$ cm, the minimum distance between two vertices 0.1 cm, $NDOF=2\times\Sigma_{tracks}(weights)-3>0.5$ where the $weights$ are the quality of the tracks associated with the vertex, at least 2, 3 or 4 tracks with $p_T>200$~MeV/$c$ in $|\eta|<2.4$ associated with the vertex and each track should have at least 2 pixel and 5 strip hits. To obtain the true number of vertices from the visible number of vertices a bin-by-bin correction factor was applied using the full simulation of the CMS detector accounting for vertex reconstruction efficiency, vertex merging and fakes.

After the corrections applied, the distributions of the fraction of pile-up events as a function of the single bunch luminosity for $n=0,...,8$ pile-up events are well fitted with a Poisson distribution. The 9 values of the cross section are fitted together to get the visible inelastic $pp$ cross section. The main sources of systematic uncertainties are the use of different datasets, change in the fit limits, the minimum distance between vertices, the vertex quality requirement and the application of an analytic method for the corrections.

The final results with at least 2, 3 or 4 charged particles with $|\eta|<2.4$ and  $p_T> 200$ MeV/$c$ of the visible inelastic $pp$ cross section are the following: 
\[ \sigma_{\rm{vis}}(\geq2 \mbox{ charged particles})=58.7\pm2.0{\rm (syst.)}\pm2.4{\rm (lumi.)} \mbox{ mb.} \]
\[ \sigma_{\rm{vis}}(\geq3 \mbox{ charged particles})=57.2\pm2.0{\rm (syst.)}\pm2.4{\rm (lumi.)} \mbox{ mb.} \]
\[ \sigma_{\rm{vis}}(\geq4 \mbox{ charged particles})=55.4\pm2.0{\rm (syst.)}\pm2.4{\rm (lumi.)} \mbox{ mb.} \]

\section{Extrapolation to the total inelastic cross section}

Additional Monte Carlo models were used for the extrapolation from the different hadron level definitions of the cross section to the total inelastic $pp$ cross section. The considered models were {\sc pythia~6}~\cite{Pythia6}, {\sc pythia~8}~\cite{Pythia8}, {\sc phojet}~\cite{Phojet}, {\sc epos~1.99}~\cite{EPOS}, {\sc sybill~2.1}~\cite{SIBYLL}, {\sc qgsjet 1} and {\sc qgsjet-ii}~\cite{QGSJETII} which use different models/tunings for the hard parton-parton and for the diffractive scattering cross sections. Every model show a similar trend for the measured hadron level cross sections but there are substancial differences in the expextations for the total inelastic $pp$ cross section.

For the extrapolation from the inelastic $pp$ cross section with $(\xi>5\times10^{-6})$ to the total inelastic $pp$ cross section all generators were considered except {\sc qgsjet 1} since this model used in cosmic-rays physics is only describing one of the hemispheres of the collision, which is not suitable for efficiency calculations when a single energy deposit in either one of the two HF calorimeters is required. The final result of the first analysis is
\[
\sigma_{\rm{inel}}=64.5\pm0.2(\rm{stat.})\pm1.1(\rm{syst.})\pm2.6(\rm{lumi.})\pm1.5(\rm{extr.})\mbox{ mb.} 
\]

The extrapolation from the visible cross section with at least 2,3 or 4 charged particles with $p_T>200$ MeV/$c$ and $|\eta|<2.4$ has been computed using only the models in agreement with the measured points (not {\sc phojet} or {\sc sybill}). The final result of the second analysis is
\[
\sigma_{\rm{inel}}=68\pm2.0(\rm{syst.})\pm2.4(\rm{lumi.})\pm4(\rm{extr.})\mbox{ mb.} 
\]

\clearpage

\begin{wrapfigure}{rt}{0.6\textwidth}
  \centering
  \includegraphics[width=0.6\textwidth]{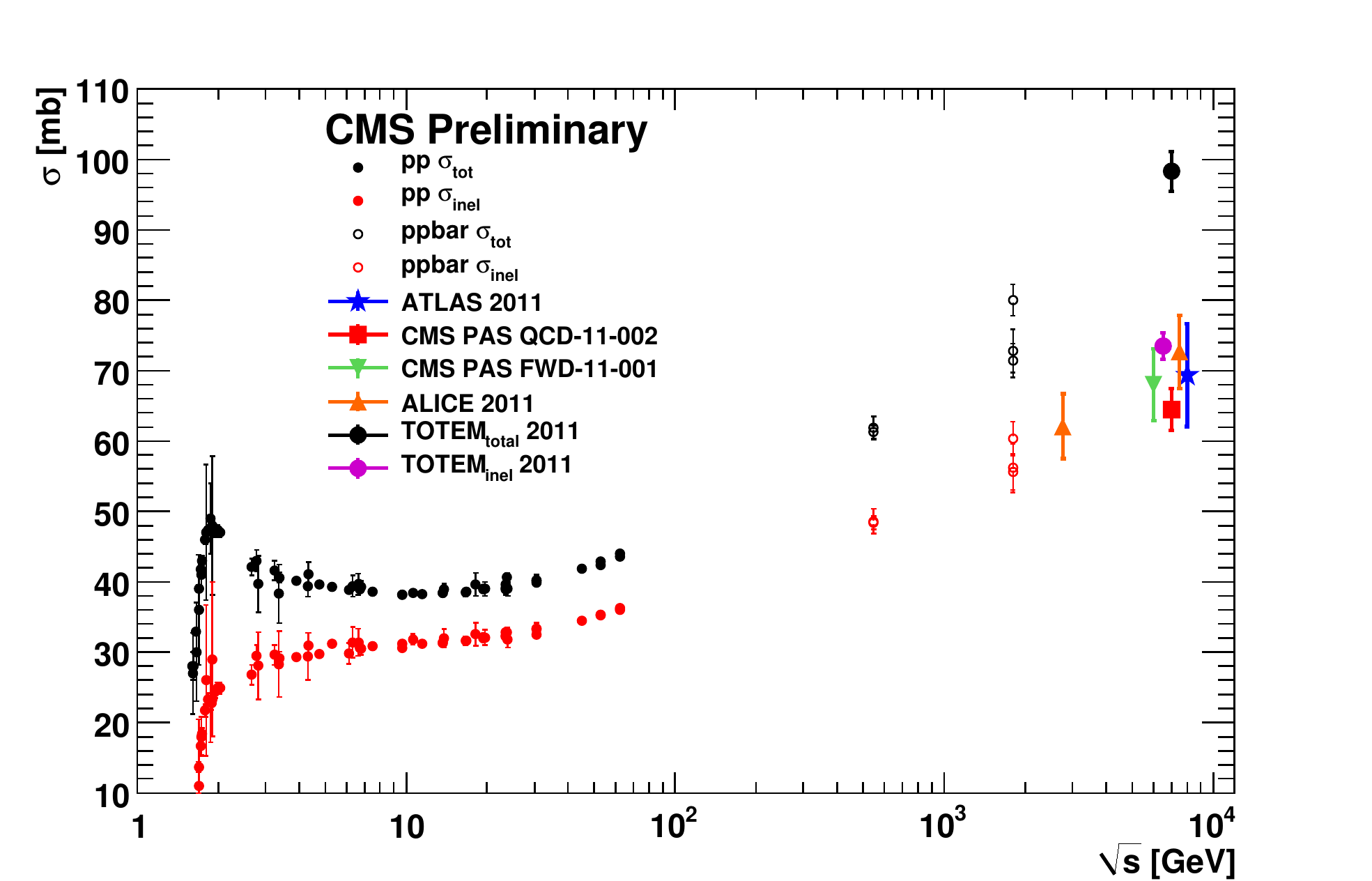}
  \caption{The total inelastic $pp$ cross section at $\sqrt{s}=7\mbox{ TeV}$ from CMS compared with ATLAS, ALICE, TOTEM and lower energy $pp$ and $p\bar{p}$ data from PDG.}
  \label{Fig:summary}
\end{wrapfigure}

\section{Summary}

CMS measured the inelastic $pp$ cross section at $\sqrt{s}=7$ TeV with two independent method using two different subdetectors. The results for the inelastic $pp$ cross section with $(\xi>5\times10^{-6})$ are in very good agreement with the result of the ATLAS Collaboration in the same kinematic range~\cite{ATLAS}.

The extrapolations to the total inelastic $pp$ cross section rely entirely on the models and their description of the low mass region. Within the large extrapolation uncertainties the results of CMS presented here are in agreement with the results from ATLAS, ALICE and TOTEM collaborations as shown in Figure~\ref{Fig:summary}.

\section*{Acknowledgements}

The author wishes to thank to the Hungarian Scientific Research Fund (K 81614) for support.


{\raggedright
\begin{footnotesize}

\end{footnotesize}
}


\end{document}